\documentclass[12pt]{iopart}

\usepackage{graphicx,color}
\usepackage{amssymb}
\usepackage{graphicx}% Include figure files
\usepackage{epsf}
\usepackage{pstricks}
\usepackage{dcolumn}% Align table columns on decimal point
\usepackage{bm}% bold math

\def\um{\mu\mbox{m}}

\def\Wcm2{\mbox{W cm}^{-2}}
\def\Wcmum2{\mbox{Wcm}^{-2}\mu\mbox{m}^{2}}

\def\Vm{\mbox{V/m}}
\def\cm3{\mbox{cm}^{-3}}

\begin{document}

\title{Dynamic Control of Laser Produced Proton Beams}

\author{S.~Kar\footnote{Electronic address: s.kar@qub.ac.uk}, K.~Markey, P.T. Simpson, B. Dromey, M. Borghesi, M. Zepf}
\address{The Queen's University of Belfast, Belfast BT7 1NN, UK}
\author{C. Bellei, S.R. Nagel, S. Kneip, L. Willingale, Z. Najmudin, K. Krushelnick\footnote{Currently at CUOS, University of
Michigan, Ann Arbor 48109, USA}}
\address{Blackett Laboratory, Imperial College, London, SW7 2BW, U.K.}
\author{J.S.~Green, P. Norreys, R.J. Clarke, D. Neely}
\address{Central Laser Facility, Rutherford Appleton Laboratory,
Chilton, OX11 0QX, U.K.}
\author{D.C.~Carroll,P. McKenna}
\address{SUPA, Department of Physics, University of Strathclyde,
Glasgow, G4 0NG, U.K.}
\author{E.L.~Clark}
\address{Department of Electronics, Technological Educational
Institute of Crete, Chania, Crete, Greece}

\date{\today}

\begin{abstract}

The emission characteristics of intense laser driven protons are
controlled using ultra-strong (of the order of $10^{9}~\Vm$)
electrostatic fields varying on a few ps timescale. The field
structures are achieved by exploiting the high potential of the
target (reaching multi-MV during the laser interaction). Suitably
shaped targets result in a reduction in the proton beam divergence,
and hence an increase in proton flux while preserving the high beam
quality. The peak focusing power and its temporal variation are
shown to depend on the target characteristics, allowing for the
collimation of the inherently highly divergent beam and the design
of achromatic electrostatic lenses.

\end{abstract}


\maketitle

\section{Introduction}

One of the most dynamic fields in arena of particle acceleration has
been the area of laser-plasma based
accelerators~\cite{ebeam_nature1,ebeam_nature2,ebeam_nature3,Clark_PRL_2000,Snavely_PRL_2000,hegelich_nature}.
This rapid development and world-wide interest is driven by the fact
that far larger accelerating gradients can be exploited than in
conventional accelerators (in excess of $10^{12}$
V/m~\cite{hegelich_nature} compared to 10's of MeV/m in conventional
designs). Consequently, laser driven accelerators hold the promise
of compact accelerating structures, which may be advantageous for a
number of applications~\cite{Borghesi_FST_2006}. After only a very
short period of development laser-accelerated proton beams
~\cite{Clark_PRL_2000,Snavely_PRL_2000} have been shown to have very
small longitudinal and transverse emittance~\cite{Borghesi_PRL_2004,
cowan_PRL_2004}. However, the excellent emittance is the result of a
very short initial burst duration and a very small virtual source
size. By extension the proton beam emerging from the target has a
broad energy spectrum ($100\%$ up to $E_{max}$) and large, energy
dependent, divergence angle (typically 40 - 60 $\deg$ depending on
laser and target parameters). The manipulation of laser generated
proton beams gives new challenges due to the high bunch charge and
short pulse nature of the beams, requiring innovative approaches to
enable beam control - recently a technique employing electric fields
triggered by a second laser pulse on a separate target, has been
used for focusing selectively a portion of the beam spectrum
~\cite{Toma_Science_2006}. However, the inherent large divergence
and energy spread can make it hard to utilise the full flux of the
proton beam for applications and indeed for further transport and
beam manipulation. Here we demonstrate a novel target configuration
which, without the need for an auxiliary laser pulse, exploits the
self-charging of the target to improve the collimation of the entire
proton beam, while conserving the characteristic high laminarity
required for radiography applications~\cite{Andy_PRL_2005}. This
approach also allows the control of chromatic properties of the beam
and creation of achromatic electrostatic lenses, by exploiting the
strong temporal variation of the target potential. Hence this
technique allows the full flux of the proton beam to be used in many
demanding applications in science, medicine and
industry~\cite{Borghesi_FST_2006}.

\begin{figure}
\begin{center}
\includegraphics[angle=0,width=0.6\textwidth]{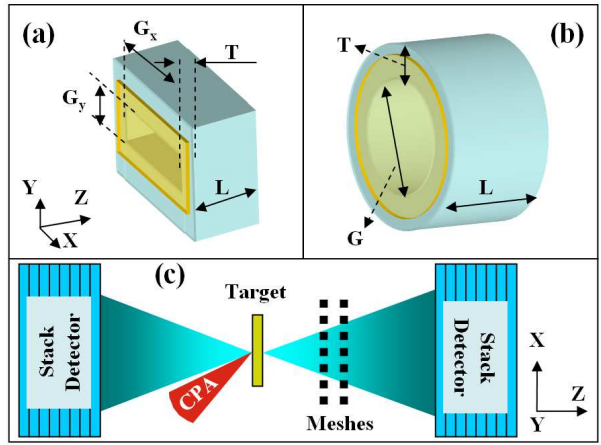}
\end{center}
\caption{Schematic of \textbf{(a)} rectangular and \textbf{(b)}
cylindrical lens-targets fielded in the experiment. \textbf{(c)}
Schematic of the experimental setup (top view).} \label{expt_setup}
\end{figure}

In high power laser interactions with solid targets, a significant
fraction of the laser energy gets transferred to the hot electron
population. Above a peak laser intensity $10^{19}~\Wcm2$,
experiments~\cite{Wharton_PRL_1998} and particle-in-cell
simulations~\cite{Wilks_IEEE_1997} have reported laser-to-electron
conversion efficiencies up to 50\%. Most of the absorbed energy is
carried by the forward moving hot electron population with an
electron spectrum that can be approximated as an exponential $dN/dE
= (N_{0}/U_{p})\times e^{-E/U_{p}}$, with a temperature of the order
of the ponderomotive potential of the incident laser
[$U_{p}=0.511\times(\sqrt{1+(a_{0}^{2}/2)}-1)$ MeV, where $a_{0}$ is
the normalized laser vector potential]. A small fraction of the hot
electron population escapes and rapidly charges the target to a
potential of the order of $U_p$ preventing the bulk of the hot
electrons from escaping. The time dependence of the target potential
is governed by the self-capacitance of the target ($C_{T}$). At a
given time t, the number of electrons that have escaped from the
target is given by $N_{es}(t) = N_{0}\times
e^{-E_{cutoff}(t)/U_{p}}$ which satisfies the equation,
$eN_{es}(t)/C_{T}(t) = 10^{-6}\times E_{cutoff}(t)$. Here $e$ is
electron charge and $E_{cutoff}$ is the cut-off energy (in MeV) of
the confined electron spectrum. Assuming the lateral velocity of the
charge wave equals to $0.75c$~\cite{Paul_PRL_2007}, the time
dependent target self-capacitance can be estimated as that of a disk
of increasing radius $C_{T}(t) \sim 8\epsilon_{0}(r_{0}+0.75ct)$,
where $\epsilon_{0}$ is permitivity of vacuum and $r_{0}$ the laser
spot size on the target. From these considerations it follows that
the target potential remains above several MV for all relevant time
scales considered here. Targets charged to similar potentials have
been shown to produce substantial transverse deflections of of
multi-MeV proton beams ~\cite{Borghesi_APL_2003}. Suitable design of
the target geometry (as shown later) therefore enables a strong
electrostatic lens to be created.

\section{Experimental Setup}

The experiment was performed at Rutherford Appleton Laboratory
employing VULCAN Petawatt laser system. The laser pulse delivered
$\sim$~300J of energy on target in ~500-600 fs FWHM duration. Using
an f/3 off-axis parabola, the laser was focussed to 8 $\um$ FWHM
spot with a peak intensity $\sim 10^{21}~\Wcm2$. Two basic shapes of
electrostatic lens geometries, rectangular and cylindrical
(Fig.~\ref{expt_setup}(a-b)) were investigated and compared to the
performance of flat Au foil targets. For all shots the laser
interacted with a 15 $\um$ thick Au foil at its centre, at an angle
of incidence of ~40 degrees from target normal. All targets were
mounted on $\sim$3 millimeter thick and $\sim$2 centimeters long
plastic stalks in order to provide a highly resistive path to the
current flowing from target to ground. Spatial and spectral profiles
of the multi-MeV proton beam emitted normally from either side of
the target were measured by employing stacks of radiochromic films
(a dosimetry detector~\cite{Borghesi_POP_2002}) as shown in
Fig.~\ref{expt_setup}(c). The dose response of the RCF detectors
were calibrated by exposing them to various known doses of protons
from a linear accelerator~\cite{Toma_private}. In some shots, Cu
sheets of suitable thickness were inserted in the stacks in order to
obtain proton spectrum via activation
measurements~\cite{Paul_PRE_2004}, complementary to those obtained
from RCF dose measurements. Multiple periodic meshes were introduced
at the rear side of the proton generating target, at various
distances from it. Multiple mesh radiographs enable virtual proton
source measurement~\cite{Borghesi_PRL_2004}, as well as to map the
proton trajectories.

\begin{figure}
\begin{center}
\includegraphics[angle=0,width=0.69\textwidth]{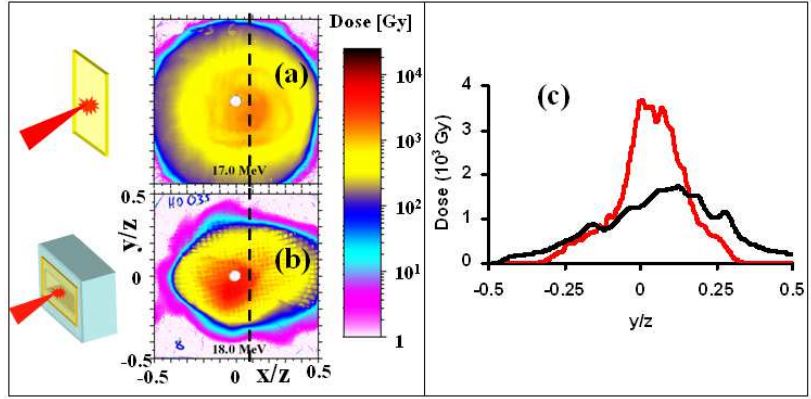}
\end{center}
\caption{Experimentally obtained proton beam (of energies ~ 17.5
MeV) spatial dose profiles for \textbf{(a)} plain foil target and
\textbf{(b)} a Al rectangular lens-target of $G_{x}$=5 mm, $G_{y}$=1
mm, L=2 mm and T=0.5 mm. Spatial scales of the images are normalized
to their distances from the proton generating foil. \textbf{(c)}
Lineouts along the dotted line in \textbf{(a)} (black) and
\textbf{(b)} (red) are shown.}\label{data}
\end{figure}

\section{Experimental results and analysis}

\begin{figure}
\begin{center}
\includegraphics[angle=0,width=0.45\textwidth]{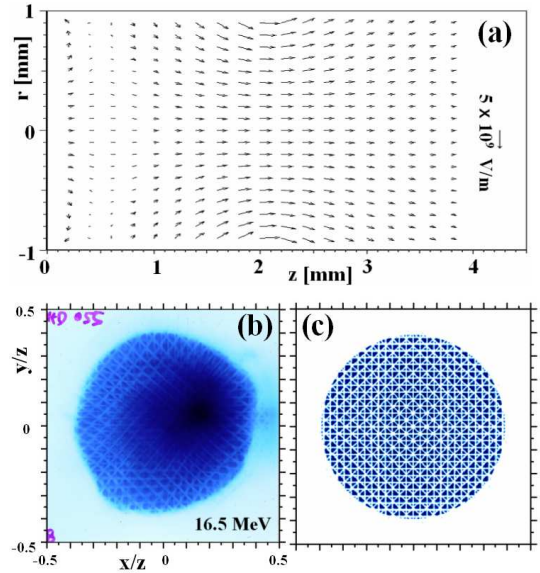}
\end{center}
\caption{\textbf{(a)} Longitudinal (across XZ plane) electric field
profile across a cylindrical Al lens-target of G=2 mm, L=2 mm
and T=0.5 mm. Experimental \textbf{(b)} and simulated \textbf{(c)}
proton spatial dose profiles (mainly due to 16.5 MeV protons)
obtained from the cylindrical lens-target. Spatial scales of the
images are normalized to their distances from the proton generating
foil.} \label{ptrace}
\end{figure}

Significant (up to a factor of two) reductions in the proton beam
divergence angle were observed in case of the lens-targets when
compared to free-standing flat foils. The angular divergence of the
low energy part (up to 25 MeV) of the spectrum was observed to be
highly reproducible ($\sim 56^{\circ}$) for the flat foil targets
(typical proton beam profile is shown in Fig.~\ref{data}(a)). In
order to ensure that the observed effect was unambiguously due to
the target geometry, rectangular lens-targets (see
Fig.~\ref{expt_setup}(b)) were employed. Fig.~\ref{data}(b) shows a
2D flux profile of the proton beam obtained from a rectangular
lens-target. Instead of the typical near-circular beam profile, an
elliptical profile (with a major to minor axes ratio of up to 2:1 as
shown in Fig.~\ref{data}(b)) is observed. Only the divergence of
protons that have propagated through the lens is affected (i.e. the
protons propagating towards the laser are not affected by the
modified target geometry). Mounting the rectangular lens on the
front surface of the proton generating foil, resulted in an
elliptical spatial profile of the proton beam propagating towards
the laser. As expected in this case, the proton beam from the rear
side of the target retains a spatial dose profile similar to that
from flat foil target. Employing lens-targets with cylindrical
symmetry (Fig.~\ref{expt_setup}(b)), resulted in a near-circular
profile of rear side protons (Fig.~\ref{ptrace}) with significant
reduction in beam divergence as compared to the flat foil case.

The spectral shape and overall dose is not affected by the addition
of a lens element to the target geometry. Spatially integrated
proton beam spectra from both plain foil and lens-targets agree to
within the typical shot to shot variations. Consequently an increase
in the proton flux commensurate with the reduction in beam size is
seen in the case of lens-targets with respect to the plain foil case
(for example, see Fig.~\ref{data}(c)). Moreover, no significant
change in the proton beam cut-off energy was found in case of the
lens-targets in comparison to the plain foil targets. This suggests
that the accelerating fields are not affected by the addition of a
lens structure. This is in good agreement with expectations since
the time it takes for the charge wave to spread to the lens far
exceeds the proton acceleration time \cite{Paul_PRL_2007}.

3D particle tracing simulations, employing
PTrace~\cite{Angelo_thesis}, were carried out in order to study the
focussing due to the electrostatic lens formed by the self-charging
of the target. The temporal evolution of the target potential is
calculated by considering the self-capacitance of a charged disk of
radius $r_{0}+0.75ct$ and an exponential electron distribution with
a temperature of $U_{P}$ as described above. The potential of the
target is assumed constant after the charge wave reaches the end of
the lens at a time time $t_{ss}$, such that
$r_{0}+0.75ct_{ss}=G/2+L$ ($G$ and $L$ are the inner diameter and
length of the cylindrical lens as shown in Fig.~\ref{expt_setup}(b).
The electric field at a given point and at a given time is obtained
by superimposing the contributions from every part of the charged
target. Test particles (protons) are launched from a point source
with a divergence and initial position taken from experimental
observations on shots with a flat foil. As shown in the
Fig.~\ref{ptrace}(a), the steady state electric field profile due to
a circular lens-target fielded in the experiment resembles that of a
conventional electrostatic \emph{Einzel} lens and therefore acts to
reduce the proton beam divergence by the strong transverse field
near the edge of the charged lens-target. Since the proton beam is
strongly divergent inside the lens the longitudinal field also
contributes to the collimation of the beam.

\begin{figure}
\begin{center}
\includegraphics[angle=0,width=0.6\textwidth]{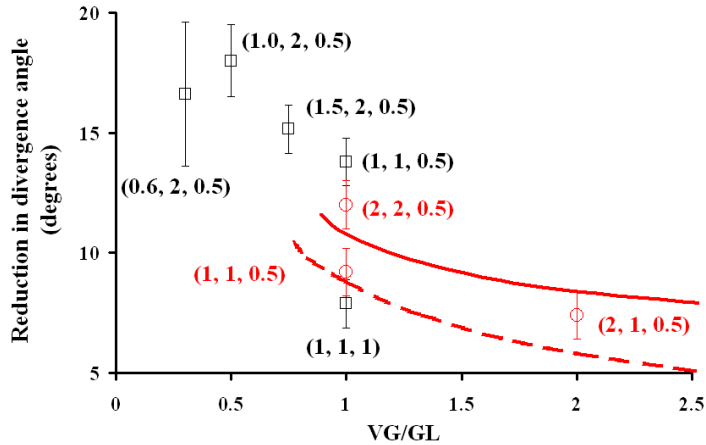}
\end{center}
\caption{Experimentally obtained data points showing reduction in
proton beam (of 17.5 MeV) divergence (along Y axis) due to
rectangular (square) and cylindrical (circle) aluminum lens-targets
of different dimensions. Dimensions (in mm) of the
rectangular(cylindrical) targets are mentioned as [$G_{y}(D), L,
T$], next to the respective data points. All rectangular
lens-targets had $G_{x}$=5 mm. Solid(dotted) lines are simulated
values for 0.5 micron thick cylindrical lens-targets of $G_{y}$
equals to 2 mm(1 mm).} \label{comparison}
\end{figure}

As shown in Fig.~\ref{ptrace}(b) and (c), the simulated proton beam
parameters at the detector plane closely match the experimentally
observed beam in both the overall dimension (divergence) as well as
the mesh magnification. Good agreement between model and observation
was obtained across different energies and for all circular
lens-targets as shown in Fig~\ref{comparison}.

Both the simulations and the experimental data show no  significant
loss of intrinsic proton beam emittance. This is demonstrated by the
clearly visible radiographs of the mesh (both experimentally as well
as in the simulations). Radiographs taken with two meshes confirm
that the that the protons follow straight line trajectories outside
the focussing field region.

From the comparison of simple Einzel-lens type configuration it is
clear that the properties of the proton beam can be manipulated
effectively. From the presented data it is clear that compensating
for the residual divergence (around 20 degrees) could be achieved
with a secondary lens target~\cite{Toma_Science_2006} and result in
a beam collimated to a diameter of a few mm size. The strong role
that the detailed geometry plays can be seen in
Fig.~\ref{comparison}, where the reduction in the proton beam
divergence for a number of lens-targets of different geometries is
shown. Collimation increases with decreasing aspect ratio $G_{y}$/L
for the rectangular lens (or G/L for the cylindrical case) and
decreases with increasing wall thickness T. These effects are
consistent with the higher fields obtained for a given potential for
smaller structures, changes in the interaction length and variations
in target capacitance respectively. Clearly, modifying the geometry
towards more complex shapes may also yield improved
collimation/focusing performance.

\begin{figure}
\begin{center}
\includegraphics[angle=0,width=0.6\textwidth]{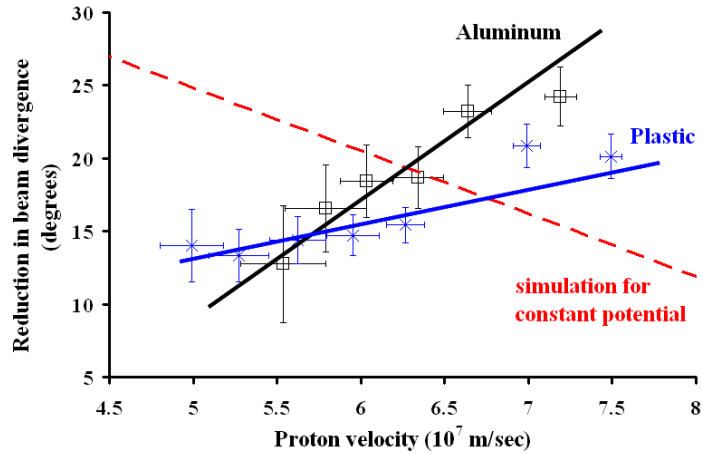}
\end{center}
\caption{Trend (solid line) of reduction in divergence over
different proton beam energy for two rectangular ($G_{x}$=5 mm,
$G_{y}$=0.6 mm, L=2 mm and T=0.5 mm) lens targets of different
materials, mentioned next to the respective trend lines. Red dashed
line illustrates chromatic behaviour of a simulated static
electrostatic lens.} \label{discharging}
\end{figure}

So far we have only discussed the behaviour of the proton beam at a
given proton energy. However, the focusing strength of an
electrostatic lens varies strongly with beam energy - i.e. it is
chromatic. Harnessing the full potential of these beams (e.g. by
focusing them to one spot or coupling the whole bunch into a
post-accelerator or phase-rotator) requires that the chromatic
behaviour be addressed. The dynamic charging and discharging of the
investigated structures allow a significant control of the chromatic
behaviour of the electrostatic lens. As the charge wave spreads at
much higher speed than the protons, complete charging up of the
target is attained much earlier than the protons leave the active
focussing field region. The protons are therefore still in the field
of the lens during the discharging period. The target discharges
primarily due to the current flowing through the stalk holding the
target. In our case, abrupt discharging of the lens target was
avoided by employing insulating stalks. Clearly the discharge time
scale will also be a function of the lens material. As shown in
Fig.~\ref{discharging} the Aluminium targets discharge (and hence
lose their focusing field) much more rapidly than the plastic
targets. For our geometry the discharge time constant for the
Aluminium lens was roughly 7.5 ps, resulting in substantial drop of
the lens potential during the particle transit. Consequently, the
slower particles see a reduced collimating field and experience a
smaller reduction in beam divergence. Note that for a constant
potential the slower protons would experience a larger drop in beam
divergence inversely proportional to the proton velocity (see
Fig.~\ref{discharging}). In the case of the plastic lens the
discharge time is clearly much longer ($\sim$15 ps). In this case
the slower reduction in lens potential almost perfectly compensates
the enhanced collimation expected at lower energies for constant
lens potential, resulting in an achromatic collimating system.
Clearly, similar approaches can be taken to achieve achromatic
focusing or a compensation of the initial chromatic dependency of
the proton beam divergence.  For example mounting the target with
much thinner (few microns) insulating wires is expected to result in
negligible discharging of the target during the relevant time
scale~\cite{Baton_HEDP_2007}, resulting in significant improvement
in lower energy proton beam collimation.

\section{Possible Advancements}

Indeed, further improvement in the beam collimation can be achieved
by suitably modified lens-target designs. For instance, reducing the
wall thickness (T) from 0.5 mm to 50$\um$ will substantially
increase the surface charge density. Similarly, the observed
increase in beam collimation with longer $L$ and shorter $G$
suggests, a conical (instead of cylindrical) lens geometry will
produce stronger focussing by guiding the beam optimally from its
source. For example, in case of a 30 degree conical lens-target with
initial diameter of 0.4 mm, L=2 mm and T=50$\um$, simulations
predict in excess of factor of three reduction of few MeV proton
beam divergence for the discussed experimental conditions. This
implies an order of magnitude increase in the proton flux which
would be promising for high energy density physics and fusion
research.

\section{Conclusions}

In conclusion, we have shown that by careful choice of target
geometry an electrostatic lens can be formed which improves the
collimation of the protons emitted from a laser irradiated foil
target. Significant reductions in the beam divergence and
commensurate increase in flux have been observed without sacrificing
the high beam quality in excellent agreement with 3D particle
tracing simulations. In addition, it is possible to offset the
chromatic nature of an electrostatic lens with the discharging of
the lens structure allowing for near-achromatic lens performance and
a route for correcting chromatic aberrations inherent in the proton
beam production process.

\section{Acknowledgements}

This work is funded by EPSRC. M.Zepf is the holder of a Royal
Society Wolfson Merit Award. Authors acknowledge supports from the
target fabrication group of RAL. SK would like to thank Dr. A.
Schiavi for the code PTRACE.

\end{document}